\documentclass[pra,floatfix,preprint,superscriptaddress,nofootinbib,eqsecnum,12pt]{revtex4}
\usepackage{hyperref}
\usepackage{epsfig}
\usepackage{bm}
\usepackage{amsmath}
\usepackage{ulem}
\input epsf

\def\HP{$(I,\Psi)$ }

\def\htk{h^{TT}_{ij}(\vec k)}
\def\arg{h_{ij}(\vec x),\chi(\vec x)}
\def\qb{\bar q}
\def\sp{ {sp}}
\def\Isp{I_{sp}[h_{ij}(\vec x),\chi(\vec x)]}
\def\sp1{h_{ij}(\vec x),\chi(\vec x)}
\def\Iv{\check{I}}
\def\Sv{\check{S}}

\def\vx{{\vec x}}

\def\ef{}

\def\c{d}

\def\be{\begin{equation}}
\def\ee{\end{equation}}

\begin{document}
\vspace{1cm} 
\title{What is  the No-Boundary Wave Function of the Universe?}

\author{J.J.~Halliwell}
\affiliation{Blackett Laboratory, Imperial College, London SW7 2BZ, UK}
\author{J. B. Hartle}
\affiliation{Department of Physics, UCSB, Santa Barbara, CA 93106, USA}
\author{T. Hertog}
\affiliation{Institute for Theoretical Physics, KU Leuven, 3001 Leuven, Belgium}

\begin{abstract}

We specify the semiclassical no-boundary wave function of the universe without relying on a functional integral of any kind. The wave function is given as a sum of specific saddle points of the dynamical theory that satisfy conditions of regularity on geometry and field and which together yield a time neutral state that is normalizable in an appropriate inner product. This specifies a predictive framework of semiclassical quantum cosmology that is adequate to make probabilistic predictions, which are in agreement with observations in simple models. 
{The use of holography to go beyond the semiclassical approximation is briefly discussed.}

\end{abstract}


\maketitle


\section{Introduction}
\label{intro}

The inference is inescapable from the physics of the last century that we live in a quantum mechanical universe.  If so, the universe must have a quantum state, which we denote by $\Psi$. A theory of that state is a necessary part of any ``final theory''  along with a theory of the dynamics which we denote by $I$. There are no predictions of any kind that do not involve both at some level. Theories of dynamics $I$ can be specified by an appropriate action as in a quantum field theory coupled to relativistic gravity. Our theory is thus \HP. This paper is concerned with the specification of the no boundary quantum state as a candidate for $\Psi$. 

The no-boundary idea was introduced in \cite{HH1983}  to understand the quantum origin of the universe's classical spacetime we observe today. The singularity theorems that Hawking and others had developed showed that our universe could not have a classical beginning with a Lorentzian geometry. Earlier work by one of us with Hawking \cite{Hartle:1976tp} on the quantum radiation from black holes had demonstrated the power of Euclidean geometry in understanding quantum effects in gravity. It was thus a natural conjecture that the universe could have a quantum beginning {corresponding, at least semiclassically, to} a Euclidean geometry \cite{Hawking1981}.

Observations suggested the universe was simpler earlier than it is now --- more homogeneous, more isotropic, more nearly in thermal equilibrium. Simplicity is a characteristic of ground states in many familiar physical systems.  So it was a natural idea that the state of the universe should be something like the cosmological analog of a  ground state.  Ground states of familiar systems are the lowest eigenstate of the system's Hamiltonian. The analog of the Hamiltonian is zero in time-reparametrization invariant theories like general relativity.  But the wave function of the ground state of many familiar systems  can also be calculated by a Euclidean functional integral  of the form $\int\exp{(-I/\hbar)}$, where $I$ is the dynamical action. It was therefore natural to conjecture that the no-boundary wave function (NBWF) could be defined by a similar integral over an appropriate contour of integration such that  it converged and satisfied the Wheeler-DeWitt (WD) equation. It was hoped that the integral would lead to a deeper connection of the wave function of the universe with quantum gravity. 

The semiclassical approximation to the wave function is logically independent of any integral representation. Properties of functional integrals can thus not be used to falsify the semiclassical NBWF. The authors know of no fundamental quantum mechanical reason why a cosmological wave function has to be defined by an integral of any particular kind.  Most wave functions in ordinary quantum mechanics are rather defined as appropriate solutions of the Schr\"odinger equation. Indeed the mathematical complexities associated with integrals over quantum spacetimes lead one to suspect there might be a simpler, more direct approach to defining cosmological wave functions in the semiclassical approximation. This is relevant also for the comparison of theories of the state of the universe with observations which is mostly done in the semiclassical approximation only.

There is a simpler approach. In this paper we define the semiclassical  NBWF {\it directly } as a collection of  appropriately regular saddle points of the action of the dynamical theory coupled to gravity. We also provide a predictive framework for semiclassical quantum cosmology that is adequate to derive probabilistic predictions for observable features of the universe from semiclassical wave functions. The collections are not arbitrary.  They  must satisfy some general conditions which include the following:  (1)  The wave function must satisfy the constraints of general relativity and the matter theory.  (2) The wave function must be consistent with the principles of the quantum framework used to make predictions. For example it must be normalizable in an appropriate Hilbert space inner product. (3)  At least at the semiclassical level the wave function must provide predictions that are consistent with observation. We will find that these principles combined with the assumptions of saddle point uniqueness and time reversal invariance are sufficient to fix the definition of the semiclassical NBWF. No integral is necessary\footnote{While we were finalising this article a paper by S. de Alwis \cite{deAlwis} appeared that has some overlap of approach.}.

With the definition of the semiclassical NBWF secure and based on this minimal set of principles, and assumptions, one can proceed to distinguish it clearly from other models of the state and discuss possible definitions beyond the semiclassical approximation.

The paper is structured as follows:  We begin in Section \ref{lingrav} with a discussion of a very simple example --- the ground state wave functional of linearized gravity. We show this can be represented by functional integrals in various different ways and  from saddle points of an appropriately defined action with no integral at all! Section \ref{princ} introduces the general semiclassical quantum framework for prediction that we employ throughout. Section \ref{scnb} defines the semiclassical NBWF. Section \ref{predictions} describes how the NBWF defined in Section \ref{scnb} makes predictions for our observations of the universe. Section \ref{conclusion} offers brief conclusions including a discussion of how holography could enable going beyond the semiclassical approximation.

\section{Linearized Gravity}
\label{lingrav}

\subsection{Linearizing Einstein  Gravity} 
\label{lingravity}

We consider near flat  metrics on ${\bf R} \times \bf T^3$ where $\bf T^3$ is the 3-box with opposite sides identified.  We can think of  the inside of the box as a model of the universe.  Nearly flat Lorentzian metrics have the form:
\be 
\label{linmetric}
ds^2 = [\eta_{\alpha\beta} + q_{\alpha\beta}(x)] dx^\alpha dx^\beta .
\ee 
Here $\eta_{\alpha\beta}$ is the Minkowski background metric in standard Cartesian coordinates $x^\alpha=(t,x^i)$. The quantity $q_{\alpha\beta}(x)$ is a small metric perturbation periodic on the opposite sides of the box.  The action for these perturbations is invariant under the gauge transformations  
\be
\label{gauge}
q_{\alpha\beta}(x) \rightarrow q_{\alpha\beta}(x) +\nabla_{(\alpha} \xi_{\beta)}
\ee
for arbitrary small periodic functions $\xi_{\alpha} (x)$. These gauge transformations implement diffeomorphism invariance at the linearized level. 

As a consequence of the gauge symmetry \eqref{gauge}, the linearized theory has four constraints.  These can be solved and the gauge can be fixed so as  to exhibit  the two true physical degrees of freedom  of quantum gravity as  the two components of a transverse-traceless metric perturbation,  {\it viz.} $q^{TT}_{ij}(x)$. This process is called {\it deparametrizing} the theory. 

The (Lorentzian) action for the true degrees of freedom is 
\be
\label{loract}
\ell^2 \Sv =\frac{1}{4} \int d^4 x [(\dot  q^{TT}_{ij})^2 - (\nabla_i q_{jk}^{TT})^2 ]
\ee
where, in $c=1$ units,  $\ell =\sqrt{16\pi\hbar G}$ s a multiple of the Planck length, a dot denotes a time derivative, and a square includes a sum over indices, viz. $(t_{ij})^2 \equiv t_{ij}t^{ij}$. We define the Euclidean action for these two degrees of freedom is
\be
\label{euact}
\ell^2 \Iv[q^{TT}_{ij}] =\frac{1}{4} \int d^4 x [(\dot q^{TT}_{ij})^2 + (\nabla_i q_{jk}^{TT})^2 ]
\ee

The action \eqref{loract} defines a 
free field theory equivalent to a collection of harmonic oscillators. This can be made explicit by expanding of $q^{TT}_{ij}(x)$ in discrete modes on ${\bf T}^3$ labeled by mode vectors $\vec k$ with frequencies $\omega_{\vec k}=|\vec k|$. 

The quantum mechanics of this system is  straightforward. 
Quantum states  are functionals of the two true degrees of freedom $h_{ij}^{TT}(\vec x)$ on any constant time slice of \eqref{linmetric}, say the one at $t=0$.   viz.
\be
\label{wvfn}
\psi=\psi[h^{TT}_{ij}(\vec x)].
\ee
There is no Wheeler-DeWitt constraint. The constraints have already been solved explicitly at the classical level. 
The ground state of linearized gravity  (GSLG) is the state where all the oscillators are in their ground state.  Explicitly
\be
\label{grnd-linear}
\Psi_0[h_{ij}^{TT}(\vec x) ] \propto \exp{\left[-\frac{1}{4 \ell^2} \sum_{\vec k} \, \omega_{\vec k} \, |h_{ij}^{TT}(\vec k)|^2\right]} .
\ee
{\ef This is indeed the ground state wave functional for linearized gravity arrived at by Hamiltonian methods \cite{Kuchar:1970mu,Hartle:1984ke} .} 
As we shall see, the ground state for this model is analogous to the NBWF for cosmology in many ways.
We now consider other ways that the GSLG could be specified.

\subsection{The semiclassical GSLG Defined by Saddle Points}
\label{saddle-def}

In the semiclassical approximation a wave function can be specified by the saddle points of the {Euclidean} action \eqref{euact} that match the argument of the wave function at $\tau=0$ and are regular on ${\bf R} \times {\bf T}^3$ for $\tau=(0,-\infty)$, {where $\tau = - i t$ denotes the Euclidean time coordinate}.
There is one such saddle point for each mode  $\vec k$  that  matches the argument of the wave function at $\tau=0$ and decays exponentially (regularity) as $ \tau \rightarrow -\infty$. The result, weighting all the modes equally, is just \eqref{grnd-linear}. In the linearized case the saddle point approximation is exact.

\subsection{GSLG Defined by a Euclidean  Integral Over Physical Degrees of Freedom}

This ground state wave function can be expressed as a functional integral over perturbations $q^{TT}_{ij}(x)$ on the range $(0,-\infty)$ that satisfy the boundary conditions above
\be
\label{linfi}
\Psi_0[h^{TT}_{ij}(\vec x) ] \propto \int \delta q^{TT} \exp{\{-\Iv[q^{TT}_{ij}(x)]}\} .
\ee
Since the action is quadratic, the integrals can be carried out explicitly and equality with \eqref{grnd-linear} verified.

\subsection{GSLG Defined by an Integral over  Geometry on a Complex Contour}
\label{GSLGcont}
In the full non-linear theory of Einstein gravity it is not possible to exhibit  the two true physical degrees of freedom explicitly as in the case of linearized gravity. The consequences of this for constructing wave functions by integrals can be illustrated in linearized gravity simply by retaining all of the metric $q_{\alpha\beta}$ in the action. The Euclidean version of the full linearized action is
\be
\label{eucact}
4\ell^2 I_2=\int d^4 x[(\nabla_\alpha \qb_{\beta\gamma})(\nabla^\alpha q^{\beta\alpha})-2(\nabla^\alpha\qb_{\alpha\beta})^2]+(\text{surface terms)}
\ee
where 
\be
\label{qb}
\qb^\alpha_{\beta} \equiv q^\alpha_\beta -\frac{1}{2}\delta^\alpha_\beta q^\gamma_\gamma.
\ee
This action exhibits the diffeomorphism invariance of the theory explicitly, but is unbounded below.
To see this consider the particular metric perturbation
$q_{\alpha\beta} =-2\delta_{\alpha\beta} \chi$. Its contribution to the action $I_2$ is 
\be
\label{chiact}
 -6 \int d^4x  (\nabla_\alpha \chi)^2 .
\ee

The full action is thus unbounded below and a functional integral of $\exp{(-I/\hbar)}$ over real values of $q_{\alpha\beta}$ would diverge. But an integral over an imaginary contour for $\chi$ will converge. This complex contour can be chosen to  implement  the ``conformal rotation'' of Gibbons, Hawking, and Perry \cite{Gibbons:1978ac}. Thus with an appropriate choice of a complex contour $\cal C$ we have  the following Euclidean functional integral for the ground state wave function of linearized gravity:
\be
\label{confact}
\Psi[q^{TT}_{\ij}(\vec x)]= \int_{\cal C} \delta q  \exp\{- I_2[q_{\alpha\beta}(x)]\},
\ee
which gives the correct ground state wave function \eqref{grnd-linear}. (For more details see \cite{Hartle:1988xv}.)

It is important to emphasize that such a complex contour is not an ad hoc choice.   It can  be {\it derived}  from the process of reparametization using  the methods described by Faddeev and Popov \cite{FP}. Deparametrization was the process of using the constraints and gauge conditions to eliminate from the action degrees of freedom other than the two physical ones $q_{ij}^{TT}$. Reparametrzation is the process of putting them back  in the integral defining ground state wave function \eqref{grnd-linear} to get a manifestly covariant expression for it. 

The reparametrization of linearized gravity was worked out in \cite{Hartle:1988xv}. The procedure involves inserting resolutions of the identity consisting of convergent Gaussian integrals over the unphysical degrees of freedom into the functional integral in \eqref{confact}. In this way an integral representation of the ground state wave function can be built with a covariant action, but with the conformal factor rotated to complex values so that the integral manifestly converges. The necessary gauge fixing machinery with Faddeev-Popov determinants is automatically included. That is how the conformal rotation can be  derived in linearized gravity. Indeed many forms of the action can be exhibited leading to convergent functional integrals.

\subsection{ Transition Amplitudes Are Not Wave Functions}
\label{wfprop}
We have described some integrals that the ground state wave function of linearized gravity \eqref{grnd-linear} {\it is} connected to.  In this section we consider an integral it {\it is not} connected to. It is not directly connected to the transition amplitudes  expressing their evolution in time. To see this we work in the Schr\"odinger picture denoting states of  a perturbation in mode $\vec k$ by $|\htk,t \rangle$. Since the modes are effectively harmonic oscillators the transition amplitudes are readily constructed. For example the transition amplitude between a state of zero field at $t=0$ to a state of different field at a later time $t$ is in $c=G=1$ units
\be
\label{prop}
\langle \htk,t |0,0\rangle \propto \exp\Bigg[ \frac{i\omega_k}{4\ell^2}  (\htk)^2 \cot{\omega_k t}\Bigg] .
\ee
Needless to say, \eqref{prop} is not the ground state wave function \eqref{grnd-linear}.

\subsection{What Do We Learn from Linearized Gravity?}
What we learn from the above analysis is that the ground state wave function of linearized gravity can be represented in several different ways. First, in the semiclassical approximation, it corresponds to a collection of regular saddle points one for each mode. Second, we learn that this collection can be represented by a functional integral in several different ways. The functional integral in \eqref{linfi} in the deparametrized theory can legitimately be called a Euclidean functional  integral because it is of the form $\int \exp(-I/\hbar)$ with $I$ real. But the representation in the parametrized theory in \eqref{confact} is not pure Euclidean all along the contour $\cal C$. Neither can it be called Lorentzian, that is, of the form $\int \exp(iS/\hbar)$ with $S$ real. Thus, there cannot be a distinction of principle about the form of a functional integral that defines the ground state wave function --- at least in linearized theory.

 \section{Semiclassical Quantum Framework}
\label{princ}

Before we turn to specifying the NBWF we first specify the general semiclassical quantum framework for prediction. We consider spatially closed cosmologies only. Wave functions\footnote{Estimating the risk of confusion as small, we use the term `wave function' even when it is represented by a functional as in \eqref{wavefn}.} representing quantum states are assumed to be functionals on the configuration space of metrics $h_{ij}(\vec x)$ and matter fields on a compact spatial three surface $\Sigma$ with 3-sphere topology.  For simplicity we generally consider only a single matter scalar field $\phi(x)$ which we write as $\chi(\vec x)$ on $\Sigma$  so that 
\be
\label{wavefn}
\Psi=\Psi[h_{ij}(\vec x), \chi(\vec x)]
\ee
where $\vec x$ stands for three coordinates on $\Sigma$. 

In the canonical framework  wave functions like \eqref{wavefn} must satisfy an operator implementation of the constraints of general relativity, in particular they must satisfy the  WD equation 
\be
\label{WDW}
H_0 \Psi=0.
\ee
where $H_0$ is an operator implementing the Hamiltonian constraint. The WD equation \eqref{WDW} will not have a unique solution. Our aim is to characterize the class of solutions that are no-boundary wave functions.

The useful outputs for a quantum theory of the universe \HP are the  probabilities for alternative histories of the universe that describe its observable properties on large scales of space and time. We therefore must specify what that quantum framework is. For a theory that aims at describing the early universe when no measurements were being made and no observers were around to make them, this cannot be familiar Copenhagen quantum mechanics. Rather we employ a more general framework suitable for closed systems like the universe that incorporates the ideas of Everett and decoherent histories quantum theory\footnote{For more background and detail on this see e.g. \cite{Hartle:1992as,Halliwell:2009rw,Halliwell:2011zz}.}.

The essential points of the quantum framework we use can be motivated by analogy with  the non-relativistic quantum mechanics of fields and particles in a closed box.  We can consider a set of alternative coarse-grained histories $\{c_\alpha\} ,  \alpha=1,2,\cdots$ of how the particles and fields evolve in the box. The individual coarse grained histories are represented by a set of class operators $C_{\alpha}$ defined by path integrals over fine-grained histories of particles and fields. Assuming the set of alternative histories  $\{c_\alpha\}$  decoheres, the probability  $p(\alpha)$ for an  individual history  $c_\alpha$ to occur is
\be
\label{probhist}
p(\alpha)=\frac{||C_\alpha |\Psi\rangle||^2}{||  \: |\Psi\rangle||^2}.
\ee
The inner product for the wave functions in the non-relativistic case is the usual $L_2$ one leading to square integrable wave functions.  Evidently $|\Psi\rangle$ has to be normalizable in that inner product. 

The important point for the present discussion is that a wave function representing a quantum state of the universe has to be normalizable to be part of the predictive framework giving consistent quantum probabilities. In the analogous construction in the gravitational case the class operators are for histories of geometry and field and the inner product is the induced inner product on superspace\footnote{See, for example, \cite{Halliwell:2009rw} and the references therein.}. Any normalizable wave function \eqref{wavefn} that satisfies the constraints is a possible candidate for the quantum state of our universe from which the probabilities for various types of Lorentzian four geometries describing our universe can be derived. We now describe the  semiclassical no-boundary wave function.

\section{The Semiclassical No-Boundary Wave Function}
\label{scnb}

In the semiclassical approximation the defining principles of Section \ref{princ} are enforced only to the first few orders in $\hbar$. This is the approximation which is closest to the classical behavior of the universe that we observe. This is the approximation in which quantum gravity is straightforward. This is the approximation that most directly supplies quantum probabilities from $(I,\Psi)$ for the classical histories of the universe that might occur. And, not surprisingly, this is the approximation in which the connection between state and probabilities has been most explored (see e.g. \cite{Hawking1983,HHH2008,Hartle:2007gi,Hartle:2010vi}). Working in this semiclassical approximation we aim at defining the  no-boundary wave function of the universe.  

\subsection{Semiclassical Wave Functions of the Universe}
\label{semisolns}

To explore the semiclassical approximation we begin by writing a general wave function in the following form
\be
\label{scwf}
\Psi[\arg]\equiv \exp\{-{\hat I}[\arg]/\hbar\}
\ee
thus defining the functional ${\hat I}$ that is an equivalent way of expressing $\Psi$. We then expand the WD equation \eqref{WDW} in powers of $\hbar$. In the leading order, the result is the classical Hamilton-Jacobi (HJ) equation for ${\hat I}[h_{ij}, \chi]$. That equation is solved if ${\hat I}$ is the action $\Isp$ of a saddle point (extremum) of the action defining the dynamical theory. Note that if $\Isp $ is a solution to the HJ equation then so is $-\Isp $ and $\Isp^*$. We will return to this below. The next order in $\hbar$ supplies a consistency condition on a prefactor to $\exp(-{ \Isp/\hbar})$.

A semiclassical wave function of the universe is therefore defined by a weighted collection of saddle points to the action $I[g,\phi]$ whose geometries have at least one spacelike boundary $\Sigma$ on which geometry and field match the arguments of the wave function $(h_{ij}(\vx),\chi({\vec x})$)  viz. 
\be
\label{collection}
\Psi[h_{ij}(\vec x),\chi(\vec x)]=\sum_{sp} \c_{sp} \exp{\{-I_{sp}[h_{ij}(\vec x),\chi(\vec x)]/\hbar\}}
\ee
where $\Isp$ is the action of the saddle point labeled by $sp$ and the $\c_{sp}$'s are suitable coefficients\footnote{Including, if necessary, beyond leading order determinants.}.

This kind of prescription provides a degree of unification of the theory of dynamics $I$ with the theory of the quantum state $\Psi$. Different dynamical theories $I$, such as different theories of quantum gravity, will generally lead to different theories of the quantum state $\Psi$.  But even assuming one dynamical theory $I$, different theories of $\Psi$ can arise from different choices for the collection of saddle points contributing.  If the collection is different the corresponding wave functions will be different. More importantly different collections will generally lead to different predictions for observations. That is because, as we show in Section \ref{predictions}, predictions for the probabilities of alternative classical histories of the universe follow simply and directly from saddle points.

\subsection{The Semiclassical No-Boundary Wave Function}
\label{seminbwf}

\begin{figure}[t]
\includegraphics[height=2.2in]{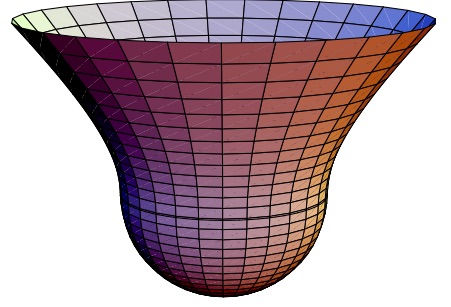}  
\caption{The iconic representation of a no-boundary saddle point in the absence of matter other than a cosmological constant. The geometry is regular and Euclidean near the South Pole and evolves across a matching surface into an expanding deSitter universe}
\label{shuttle}
\end{figure}

A semiclassical no-boundary wave function is defined by a weighted collection of saddle points (extrema) of the action $I[g,\phi]$ on a four-disk that match $(h_{ij},\chi)$ on its only boundary and are {\it otherwise regular inside}. Regularity of saddle points is what singles out the semiclassical NBWF from other semiclassical wave functions of the universe. These saddle point geometries are generally complex. Thus the NBWF captures Hawking's insight that although the universe could not begin with a regular Lorentzian geometry, it could begin with a non-Lorentzian regular geometry.  
In a rough but intuitive sense, semiclassical NBWF's are the simplest class of candidates for the quantum state of the universe defined by collections of saddle points. There are many more ways for saddle points to be irregular that to be regular. One possible geometric representation of a no-boundary saddle point in a model with a positive cosmological constant is shown in Fig. \ref{shuttle}.

Explicit calculations in minisuperspace models (see e.g. \cite{Halliwell1988,Halliwell1989,Halliwell1990,HHH2008,DiazDorronsoro:2018wro}) suggest that for a given $(h_{ij}(\vec x),\chi(\vec x))$ the actions of a no-boundary saddle point are essentially unique up to a change of sign and complex conjugation. Assuming this uniqueness, the construction of the semiclassical no boundary wave function is straightforward.

The principles of Section \ref{princ} limit which saddle points contribute to the sum \eqref{collection}. In particular the wave function must be normalizable in order for it to fit in the predictive framework sketched in Section \ref{princ} that is employed to derive probabilities for observables. { As noted above,} if $I_{sp}$ is a saddle point than $-I_{sp}$ is also. One might be tempted to include both saddle points in the sum \eqref{collection}. However these saddle points have very different physical consequences especially if the configuration space includes fluctuations away from homogeneity and isotropy. If the fluctuations are damped in one case they will be anti-damped in the other risking violating the normalization principle. One expects only the saddle points with damped fluctuations yield a normalizable wave functions  \cite{HarHal1990,Halliwell1990} and this is borne out by explicit calculations in minisuperspace models (e.g. \cite{DiazDorronsoro:2018wro}). 

When the saddle points with actions $I_{sp}$ and $I_{sp}^*$ are included with equal weight the resulting wave function will be real and time reversal invariant in a sense appropriate for gravitational physics\footnote{In non-relativistic quantum mechanics if $\psi(x,t)$ is a solution of the Schr\"odinger equation then $\psi^*(x,-t)$ is also a solution of the time-reversed Schr\"odinger equation. Then it  is consistent with dynamical theories that are also time reversal invariant. Thus, a wave function is time reversal invariant if it is real. The same result can be seen to hold for closed cosmologies in quantum general relativity using a representation where the wave function depends on the conformal three-metric ${\tilde h}_{ij}$, the field $\chi(\vec x)$, and the extrinsic curvature scalar $K(\vec x)$ which plays the role of time.}. The wave function can then be said to be invariant under C, P and T separately \cite{Hawking:1985af} consistent with dynamical theories which generally display the same symmetries. 

But what if there is not a unique saddle point for given $(\arg)$? Then there could be many NBWFs constructed from different collections of NB saddle points\footnote{This was discussed in the context of integral representations of the NBWF where different contours in integration could single out different saddle points contributing in the steepest descents approximation to the integral \cite{HarHal1990}.}.

\subsection{Classical Lorentzian Histories of the Universe from Semiclassical Saddle Points}
\label{classhist}

As already mentioned, most of what we observe of our universe on large scales are properties of its classical history --- for example,  its expansion history, the CMB, the evolution of density fluctuations into the large scale distribution of galaxies, etc.  A model of the semiclassical wave function of the universe is thus most directly tested by the probabilities it predictrs for 
the classical cosmological history of geometry and field that we observe.

Classical behavior is not a given in quantum mechanics. It is a matter of quantum probabilities.  A quantum system behaves classically when, in a particular patch of configuration space, and with a suitable set of coarse-grained alternative histories, the probability is high for histories exhibiting correlations in time by deterministic classical laws (e.g. \cite{HHH2008}). Coarse-graining is essential for classicality. For instance in the quantum mechanics of a single particle we expect classical behavior to emerge only when its position and momentum are followed to coarse-grained intervals consistent with the uncertainty principle and then only when these are followed,  not at each and every time, but only at a series of times. Similarly in eternal inflation the evolution of the universe on the largest scales is dominated by stochastic quantum effects and we expect classical behavior to emerge only after coarse graining over the fluctuations on those scales \cite{Hartle:2010dq}.

A semiclassical wave function is tested first by whether it predicts classical spacetime and matter evolution in patches of configuration space at all, and then whether the probabilities of these are significant for those with the spacetimes we observe. The saddle point geometries and fields featuring in NBWF's will generally be complex so their action has both real and imaginary parts. For a given saddle point we have
\be
\label{realimag}
I[\arg] \equiv I_{R}[\arg] - iS[\arg] .
\ee
The resulting wave function then has a WKB form, viz.
\be
\label{WkBform}
\Psi[h_{ij} (\vx),\chi(\vx)] \propto \exp \left( \{-I_{R}[\arg] +iS[\arg]\}/\hbar\right) .
\ee
where both $I_R$ and $S$ are real. Analogy with WKB in non-relativistic quantum mechanics suggests that when $S$ varies rapidly compared to $I_R$ (the classicality condition) this wave function predicts an ensemble of appropriately coarse grained\footnote{The amount of coarse graining required for classicality can vary from the minimum consistent with the uncertainty principle all the way up to large regions of the universe in the case of eternal inflation.}   classical histories of the universe which are integral curves of $S$. By `integral curve' we mean a solution of the Hamilton-Jacobi relations between the momenta conjugate to $h_{ij}$ and $\chi$ and the functional derivatives of the action $S$. Continuing the analogy we expect probabilities for these histories  to be proportional  to $\exp{\{-2I_R [\arg]/\hbar}\}$. The WD equation shows that this is constant along classical histories to lowest relevant order in $\hbar$. Of course the predicted classical histories need not be complete. One expects the above conditions for classical behavior to be predicted from the wave function not to hold inside black holes and in the early universe. This is indeed born out by explicit calculations of the original semiclassical NBWF in simple models (e.g. \cite{HHH2008,Hartle:2010dq}).

\section{Predictions for Observations}
\label{predictions}
In this section  we briefly discuss some of the key predictions of the semiclassical NBWF. We concentrate on the original NBWF \cite{HH1983} defined here semiclassically as a particular sum over saddle points which together yield a real wave function. For more details we refer the reader to \cite{Hartle:2007gi,HHH2008,Hartle:2010dq,Hartle:2010vi,Hertog:2013mra,Hertog:2015zwh}

The most striking prediction of the semiclassical NBWF is that it says our classical history emerged from a period of inflation. Cosmic inflation is the mechanism in the NBWF through which our universe became classical. In a given dynamical theory, the NBWF thus selects those regions of the scalar potential where the slow roll conditions for inflation hold. Intuitively this is because the regularity of the no-boundary saddle points requires the universe to be dominated by potential rather than gradient energy at early times. 

In dynamical models where the scalar potential has more than one slow roll region the relative probabilities of the different kinds of classical histories in the NBWF implies a relative weighting of the different models of inflation contained in the potential, which in turn yields a prior over various observables for use in comparing theory with observation. We are especially interested in such prior probabilities for local observables connected to the CMB. The NBWF predicts the usual probabilities for nearly Gaussian scalar and tensor perturbations around each inflationary background in its ensemble. Their statistical properties are specified by the shape of the potential patch probed by the background as usual. If the NBWF prior is sharply peaked around inflationary backgrounds associated with a particular region of the potential then the theory predicts the observed CMB perturbation spectrum should exhibit the features characteristic of the potential in that region.

To determine the region of a given potential that dominates the NBWF prior it is important to take in account our observational situation. The NBWF by itself is heavily biased towards histories of the universe with a low amount of inflation. However we do not observe the entire universe. Instead our observations are limited to a small patch mostly along part of our past light cone. Probabilities for local observations are conditional probabilities weighted by the volume of a surface of constant measured density, to account for the different possible locations of our past light cone. This transforms the probability distribution for the amount of inflation and leads to the prediction that our universe emerged from a region of the potential where the conditions for eternal inflation hold. Combined with the general tendency of the NBWF to favor inflation at low values of the potential \cite{HHH2008,Hartle:2007gi} this leads to the prediction that our observed classical universe emerged from the lowest exit of eternal inflation, e.g. near a broad maximum of the scalar potential.

\section{Conclusion}
\label{conclusion}

We have given a unique formulation of the semiclassical NBWF directly in terms of a collection of saddle points that satisfy a specific minimal set of criteria together with a uniqueness assumption. This formulation disentangles the status and predictions of the semiclassical NBWF from its representation as a functional integral. Beyond its obvious geometric underpinnings we advance the requirement that the saddle point NBWF be normalizable in a suitable inner product. This specifies the predictive framework of semiclassical quantum cosmology needed  to derive probabilistic predictions for features of our observed universe. 

The formulation of the NBWF given here clarifies the comparison between the NBWF and alternative models of initial conditions in semiclassical quantum cosmology such as the tunneling wave function and the purely Lorentzian functional integral approach. In its original form \cite{Vilenkin1984} the tunneling wave function involves regular no-boundary saddle points, but attempted functional integral implementations of it \cite{HarHal1990,Halliwell1988} predict undamped inhomogeneous fluctuations. The resulting wave function is likely to be non-normalizable and therefore not a candidate NBWF even though it involves no-boundary saddle points. In a recent, modified path-integral form \cite{Vilenkin:2018dch}, the tunnelling wave function predicts fluctuations are damped but this comes at the expense of the defining saddle points being singular. Hence again this does not correspond to a viable semiclassical NBWF. The Lorentzian path integral approach pursued originally in \cite{Teitelboim:1983fi} and more recently in \cite{FLT1,FLT2} does not yield a solution of the Wheeler-DeWitt equation but rather a Green's function. Moreover it is dominated by saddle points that differ from those specifying the semiclassical NBWF \cite{DiazDorronsoro2017,DiazDorronsoro:2018wro} and which yield fluctuation wave functions that imply fluctuations are not suppressed \cite{HarHal1990,Halliwell1988,FLT2}. In its current form this therefore fails to provide a predictive framework of semiclassical quantum cosmology.

Finally the form of the NBWF we put forward naturally connects to new routes towards a completion of the theory. Holographic cosmology in particular indicates that the no-boundary wave function as we know it need not have a fundamental representation as a gravitational path integral but rather emerges as an approximation to the partition functions of dual (Euclidean) field theories defined directly on the final boundary \cite{Maldacena2002,Hertog2011,Hartle:2012tv}. The arguments of the wave function in this holographic approach source deformations of the dual theory. The dependence of the partition functions on the values of the sources specifies a holographic `no-boundary' measure for cosmology. Recent explicit calculations of partition functions, mostly performed in the context of vector toy models, exhibit a remarkable qualitative agreement with the predictions of the semiclassical NBWF \cite{McFadden2009,Anninos2012,Hawking:2017wrd}. Of course this does not exclude the existence of a convenient path integral representation that organizes the saddle points of the semiclassical NBWF, with a contour specified by holography.

To summarize, a theory of a quantum state is a necessary part of any final theory of the universe. The no-boundary quantum state of the universe is defined and successfully predictive in the semiclassical approximation to the quantum cosmological predictive framework we have described. It provides a model of how theory is compared with observation in any state of the universe in quantum cosmology.  It also exhibits a promising connection with contemporary theories of quantum gravity with which it may be extended beyond the semiclassical approximation to new realms of prediction and test.

\acknowledgements
We thank Oliver Janssen for stimulating discussions. This work was supported in part by the European Research Council grant no. ERC-2013-CoG 616732 HoloQosmos, and by the US National Science Foundation under grant PHY-18-028105. 

\bibliographystyle{klebphys2}
\bibliography{references2}

\providecommand{\href}[2]{#2}\begingroup\raggedright\begin{thebibliography}{10}

\bibitem{HH1983}
{\sc J.~B. Hartle} and {\sc S.~W. Hawking}, ``{Wave Function of the
  Universe},''
\href{http://dx.doi.org/10.1103/PhysRevD.28.2960}{{\em Phys. Rev.} {\bfseries
  D28} (1983) 2960--2975}.

\bibitem{Hartle:1976tp}
{\sc J.~B. Hartle} and {\sc S.~W. Hawking}, ``{Path Integral Derivation of
  Black Hole Radiance},''
\href{http://dx.doi.org/10.1103/PhysRevD.13.2188}{{\em Phys. Rev.} {\bfseries
  D13} (1976) 2188--2203}.

\bibitem{Hawking1981}
{\sc S.~W. Hawking}, ``{The Boundary Conditions of the Universe},''
{\em Pontif. Acad. Sci. Scr. Varia} {\bfseries 48} (1982) 563--574.

\bibitem{deAlwis}
{\sc S.~P. de~Alwis}, ``The wave function of the universe and cmb
  fluctuations,''
\href{http://arxiv.org/abs/1811.12892}{{\ttfamily arXiv:1811.12892 [hep-th]}}.

\bibitem{Kuchar:1970mu}
{\sc K.~Kuchar}, ``{Ground state functional of the linearized gravitational
  field},''
\href{http://dx.doi.org/10.1063/1.1665133}{{\em J. Math. Phys.} {\bfseries 11}
  (1970) 3322--3334}.

\bibitem{Hartle:1984ke}
{\sc J.~B. Hartle}, ``{Ground state wave function of linearized gravity},''
\href{http://dx.doi.org/10.1103/PhysRevD.29.2730}{{\em Phys. Rev.} {\bfseries
  D29} (1984) 2730--2737}.

\bibitem{Gibbons:1978ac}
{\sc G.~W. Gibbons}, {\sc S.~W. Hawking}, and {\sc M.~J. Perry}, ``{Path
  Integrals and the Indefiniteness of the Gravitational Action},''
\href{http://dx.doi.org/10.1016/0550-3213(78)90161-X}{{\em Nucl. Phys.}
  {\bfseries B138} (1978) 141--150}.

\bibitem{Hartle:1988xv}
{\sc J.~B. Hartle} and {\sc K.~Schleich}, ``{The conformal rotation in
  linearized gravity},'' in {\em Quantum Field Theory and Quantum Statistics:
  Essays in Honour of the Sixtieth Birthday of E.S.Fradkin}, {\sc
  G.~I.A.Batalin} and {\sc C.J.Isham}, eds.
\newblock Hilger, Bristol,
1987.
\newblock

\bibitem{FP}
{\sc L.~D. Faddeev} and {\sc V.~N. Popov}, ``Covariant quantization of the
  gravitational field,'' {\em Soviet Physics Uspekhi} {\bfseries 16} no.~6,
  (1974) 777.

\bibitem{Hartle:1992as}
{\sc J.~Hartle}, ``{Space-time quantum mechanics and the quantum mechanics of
  space-time},'' in {\em {Gravitation and quantizations, Proc. 57th Les Houches
  Summer School in Theoretical Physics}}, pp.~285--480.
\newblock 1992.
\newblock
\href{http://arxiv.org/abs/gr-qc/9304006}{{\ttfamily arXiv:gr-qc/9304006
  [gr-qc]}}.
\newblock

\bibitem{Halliwell:2009rw}
{\sc J.~J. Halliwell}, ``{Probabilities in Quantum Cosmological Models: A
  Decoherent Histories Analysis Using a Complex Potential},''
  \href{http://dx.doi.org/10.1103/PhysRevD.80.124032}{{\em Phys. Rev.}
  {\bfseries D80} (2009) 124032},
\href{http://arxiv.org/abs/0909.2597}{{\ttfamily arXiv:0909.2597 [gr-qc]}}.

\bibitem{Halliwell:2011zz}
{\sc J.~J. Halliwell}, ``{Decoherent histories analysis of minisuperspace
  quantum cosmology},''
  \href{http://dx.doi.org/10.1088/1742-6596/306/1/012023}{{\em J. Phys. Conf.
  Ser.} {\bfseries 306} (2011) 012023},
\href{http://arxiv.org/abs/1108.5991}{{\ttfamily arXiv:1108.5991 [gr-qc]}}.

\bibitem{Hawking1983}
{\sc S.~W. Hawking}, ``{The Quantum State of the Universe},''
\href{http://dx.doi.org/10.1016/0550-3213(84)90093-2}{{\em Nucl. Phys.}
  {\bfseries B239} (1984) 257}.

\bibitem{HHH2008}
{\sc J.~B. Hartle}, {\sc S.~W. Hawking}, and {\sc T.~Hertog}, ``{The Classical
  Universes of the No-Boundary Quantum State},''
  \href{http://dx.doi.org/10.1103/PhysRevD.77.123537}{{\em Phys. Rev.}
  {\bfseries D77} (2008) 123537},
\href{http://arxiv.org/abs/0803.1663}{{\ttfamily arXiv:0803.1663 [hep-th]}}.

\bibitem{Hartle:2007gi}
{\sc J.~B. Hartle}, {\sc S.~W. Hawking}, and {\sc T.~Hertog}, ``{No-Boundary
  Measure of the Universe},''
  \href{http://dx.doi.org/10.1103/PhysRevLett.100.201301}{{\em Phys. Rev.
  Lett.} {\bfseries 100} (2008) 201301},
\href{http://arxiv.org/abs/0711.4630}{{\ttfamily arXiv:0711.4630 [hep-th]}}.

\bibitem{Hartle:2010vi}
{\sc J.~B. Hartle}, {\sc S.~W. Hawking}, and {\sc T.~Hertog}, ``{The
  No-Boundary Measure in the Regime of Eternal Inflation},''
  \href{http://dx.doi.org/10.1103/PhysRevD.82.063510}{{\em Phys. Rev.}
  {\bfseries D82} (2010) 063510},
\href{http://arxiv.org/abs/1001.0262}{{\ttfamily arXiv:1001.0262 [hep-th]}}.

\bibitem{Halliwell1988}
{\sc J.~J. Halliwell} and {\sc J.~Louko}, ``{Steepest Descent Contours in the
  Path Integral Approach to Quantum Cosmology. 1. The de Sitter Minisuperspace
  Model},''
\href{http://dx.doi.org/10.1103/PhysRevD.39.2206}{{\em Phys. Rev.} {\bfseries
  D39} (1989) 2206}.

\bibitem{Halliwell1989}
{\sc J.~J. Halliwell} and {\sc J.~Louko}, ``{Steepest Descent Contours in the
  Path Integral Approach to Quantum Cosmology. 2. Microsuperspace},''
\href{http://dx.doi.org/10.1103/PhysRevD.40.1868}{{\em Phys. Rev.} {\bfseries
  D40} (1989) 1868}.

\bibitem{Halliwell1990}
{\sc J.~J. Halliwell} and {\sc J.~Louko}, ``{Steepest Descent Contours in the
  Path Integral Approach to Quantum Cosmology. 3. A General Method With
  Applications to Anisotropic Minisuperspace Models},''
\href{http://dx.doi.org/10.1103/PhysRevD.42.3997}{{\em Phys. Rev.} {\bfseries
  D42} (1990) 3997--4031}.

\bibitem{DiazDorronsoro:2018wro}
{\sc J.~Diaz~Dorronsoro}, {\sc J.~J. Halliwell}, {\sc J.~B. Hartle}, {\sc
  T.~Hertog}, {\sc O.~Janssen}, and {\sc Y.~Vreys}, ``{Damped perturbations in
  the no-boundary state},''
  \href{http://dx.doi.org/10.1103/PhysRevLett.121.081302}{{\em Phys. Rev.
  Lett.} {\bfseries 121} no.~8, (2018) 081302},
\href{http://arxiv.org/abs/1804.01102}{{\ttfamily arXiv:1804.01102 [gr-qc]}}.

\bibitem{HarHal1990}
{\sc J.~J. Halliwell} and {\sc J.~B. Hartle}, ``{Integration Contours for the
  No-Boundary Wave Function of the Universe},''
  \href{http://dx.doi.org/10.1103/PhysRevD.41.1815}{{\em Phys. Rev. D}
  {\bfseries 41} (Mar, 1990) 1815--1834}.

\bibitem{Hawking:1985af}
{\sc S.~W. Hawking}, ``{The Arrow of Time in Cosmology},''
  \href{http://dx.doi.org/10.1103/PhysRevD.32.2489}{{\em Phys. Rev.} {\bfseries
  D32} (1985) 2489}.
[Adv. Ser. Astrophys. Cosmol.3,308(1987)].

\bibitem{Hartle:2010dq}
{\sc J.~Hartle}, {\sc S.~W. Hawking}, and {\sc T.~Hertog}, ``{Local Observation
  in Eternal inflation},''
  \href{http://dx.doi.org/10.1103/PhysRevLett.106.141302}{{\em Phys. Rev.
  Lett.} {\bfseries 106} (2011) 141302},
\href{http://arxiv.org/abs/1009.2525}{{\ttfamily arXiv:1009.2525 [hep-th]}}.

\bibitem{Hertog:2013mra}
{\sc T.~Hertog}, ``{Predicting a Prior for Planck},''
  \href{http://dx.doi.org/10.1088/1475-7516/2014/02/043}{{\em JCAP} {\bfseries
  1402} (2014) 043},
\href{http://arxiv.org/abs/1305.6135}{{\ttfamily arXiv:1305.6135
  [astro-ph.CO]}}.

\bibitem{Hertog:2015zwh}
{\sc T.~Hertog} and {\sc O.~Janssen}, ``{Sharp Predictions from Eternal
  Inflation Patches in D-brane Inflation},''
  \href{http://dx.doi.org/10.1088/1475-7516/2017/04/011}{{\em JCAP} {\bfseries
  1704} no.~04, (2017) 011},
\href{http://arxiv.org/abs/1512.02722}{{\ttfamily arXiv:1512.02722
  [astro-ph.CO]}}.

\bibitem{Vilenkin1984}
{\sc A.~Vilenkin}, ``{Quantum Creation of Universes},''
\href{http://dx.doi.org/10.1103/PhysRevD.30.509}{{\em Phys. Rev.} {\bfseries
  D30} (1984) 509--511}.

\bibitem{Vilenkin:2018dch}
{\sc A.~Vilenkin} and {\sc M.~Yamada}, ``{Tunneling wave function of the
  universe},'' \href{http://dx.doi.org/10.1103/PhysRevD.98.066003}{{\em Phys.
  Rev.} {\bfseries D98} no.~6, (2018) 066003},
\href{http://arxiv.org/abs/1808.02032}{{\ttfamily arXiv:1808.02032 [gr-qc]}}.

\bibitem{Teitelboim:1983fi}
{\sc C.~Teitelboim}, ``{Quantum Mechanics of the Gravitational Field in
  Asymptotically Flat Space},''
\href{http://dx.doi.org/10.1103/PhysRevD.28.310}{{\em Phys. Rev.} {\bfseries
  D28} (1983) 310}.

\bibitem{FLT1}
{\sc J.~Feldbrugge}, {\sc J.-L. Lehners}, and {\sc N.~Turok}, ``{Lorentzian
  Quantum Cosmology},''
  \href{http://dx.doi.org/10.1103/PhysRevD.95.103508}{{\em Phys. Rev.}
  {\bfseries D95} no.~10, (2017) 103508},
\href{http://arxiv.org/abs/1703.02076}{{\ttfamily arXiv:1703.02076 [hep-th]}}.

\bibitem{FLT2}
{\sc J.~Feldbrugge}, {\sc J.-L. Lehners}, and {\sc N.~Turok}, ``{No smooth
  beginning for spacetime},''
\href{http://dx.doi.org/10.1103/PhysRevLett.119.171301}{{\em Phys. Rev. Lett.}
  {\bfseries 119} no.~17, (2017) 171301}.

\bibitem{DiazDorronsoro2017}
{\sc J.~Diaz~Dorronsoro}, {\sc J.~J. Halliwell}, {\sc J.~B. Hartle}, {\sc
  T.~Hertog}, and {\sc O.~Janssen}, ``{Real no-boundary wave function in
  Lorentzian quantum cosmology},''
  \href{http://dx.doi.org/10.1103/PhysRevD.96.043505}{{\em Phys. Rev.}
  {\bfseries D96} no.~4, (2017) 043505},
\href{http://arxiv.org/abs/1705.05340}{{\ttfamily arXiv:1705.05340 [gr-qc]}}.

\bibitem{Maldacena2002}
{\sc J.~M. Maldacena}, ``{Non-Gaussian Features of Primordial Fluctuations in
  Single Field Inflationary Models},''
  \href{http://dx.doi.org/10.1088/1126-6708/2003/05/013}{{\em JHEP} {\bfseries
  0305} (2003) 013},
\href{http://arxiv.org/abs/astro-ph/0210603}{{\ttfamily arXiv:astro-ph/0210603
  [astro-ph]}}.

\bibitem{Hertog2011}
{\sc T.~Hertog} and {\sc J.~Hartle}, ``{Holographic No-Boundary Measure},''
  {\em JHEP} {\bfseries 1205} (2012) 095,
\href{http://arxiv.org/abs/1111.6090}{{\ttfamily arXiv:1111.6090 [hep-th]}}.

\bibitem{Hartle:2012tv}
{\sc J.~B. Hartle}, {\sc S.~W. Hawking}, and {\sc T.~Hertog}, ``{Quantum
  Probabilities for Inflation from Holography},''
  \href{http://dx.doi.org/10.1088/1475-7516/2014/01/015}{{\em JCAP} {\bfseries
  1401} no.~01, (2014) 015},
\href{http://arxiv.org/abs/1207.6653}{{\ttfamily arXiv:1207.6653 [hep-th]}}.

\bibitem{McFadden2009}
{\sc P.~McFadden} and {\sc K.~Skenderis}, ``{Holography for Cosmology},''
  \href{http://dx.doi.org/10.1103/PhysRevD.81.021301}{{\em Phys.Rev.}
  {\bfseries D81} (2010) 021301},
\href{http://arxiv.org/abs/0907.5542}{{\ttfamily arXiv:0907.5542 [hep-th]}}.

\bibitem{Anninos2012}
{\sc D.~Anninos}, {\sc F.~Denef}, and {\sc D.~Harlow}, ``{Wave Function of
  Vasiliev's Universe: A Few Slices Thereof},''
  \href{http://dx.doi.org/10.1103/PhysRevD.88.084049}{{\em Phys.Rev.}
  {\bfseries D88} no.~8, (2013) 084049},
\href{http://arxiv.org/abs/1207.5517}{{\ttfamily arXiv:1207.5517 [hep-th]}}.

\bibitem{Hawking:2017wrd}
{\sc S.~W. Hawking} and {\sc T.~Hertog}, ``{A Smooth Exit from Eternal
  Inflation?},'' \href{http://dx.doi.org/10.1007/JHEP04(2018)147}{{\em JHEP}
  {\bfseries 04} (2018) 147},
\href{http://arxiv.org/abs/1707.07702}{{\ttfamily arXiv:1707.07702 [hep-th]}}.

\end{thebibliography}\endgroup

\end{document}